\documentclass[onecolumn,preprintnumbers,eqsecnum,superscriptaddress,11pt]{revtex4}
\usepackage{amsfonts}
\usepackage{amssymb}
\usepackage{amsmath}
\usepackage{epsf}
\usepackage{graphicx}
\usepackage{dcolumn}
\usepackage{bm}
\usepackage{CJK}
\usepackage{color}
\usepackage[colorlinks=true,
            citecolor=blue,
            urlcolor=blue,
            filecolor=black,
            linktocpage=true,
            linkcolor=blue,,
             ]{hyperref}

\def\nn{{\nonumber}}

\newcommand{\beq}{\begin{equation}}
\newcommand{\eeq}{\end{equation}}
\newcommand{\beqs}{\begin{eqnarray}}
\newcommand{\eeqs}{\end{eqnarray}}

\newcommand{\be}{\begin{equation}}
\newcommand{\ee}{\end{equation}}
\newcommand{\bea}{\begin{eqnarray}}
\newcommand{\eea}{\end{eqnarray}}

\def\tJ{\tilde{J}}

\def\tF{\tilde{F}}
\def\tA{\tilde{A}}
\def\tildeF{\tilde F}

\def\tsigma{\tilde{\sigma}}

\def\Chone{C_{h1}}
\def\Chtwo{C_{h2}}
\def\Ckone{C_{k1}}
\def\Cktwo{C_{k2}}
\def\Cavone{C_{a1}}
\def\Cavtwo{C_{a2}}
\def\Cthree{C_{j1}}
\def\Cfour{C_{j2}}
\def\CTHREE{C_{j1}}
\def\CFOUR{C_{j2}}
\def\GN{G_{N}}

\def\qq{\qquad}
\def\[{\left[}
\def\]{\right]}
\def\({\left(}
\def\){\right)}

\begin{document}

\title{Holographic Charged Fluid with Anomalous Current at Finite Cutoff Surface in Einstein-Maxwell Gravity}

\author{{
{Xiaojian Bai,$^{1,2}{}^\star$\email{baixj@sogang.ac.kr}}
{Ya-Peng Hu,$^{1,2,3,4}{}^\dagger$}\email{huyp@nuaa.edu.cn}
{Bum-Hoon Lee,$^{2}{}^\ddagger$}\email{bhl@sogang.ac.kr}
and {Yun-Long Zhang$^{5}{}^\S$}\email{zhangyl@itp.ac.cn}\\[0.4cm]
\it\small{$^{1}$College of Science, Nanjing University of Aeronautics and Astronautics, Nanjing 211106, China.\\[.5em]
\it$^{2}$Center for Quantum Spacetime, Sogang University, Seoul 121-742, Korea.\\[.5em]
\it$^{3}$Center for High-Energy Physics, Peking University, Beijing 100871, China.\\[.5em]
\it$^{4}$INPAC, Department of Physics, and Shanghai Key Lab for Particle Physics and Cosmology, Shanghai Jiao Tong University, Shanghai 200240, China. \\[.5em]
\it$^{5}$State Key Laboratory of Theoretical Physics, Institute of Theoretical Physics, \\
Chinese Academy of Sciences, Beijing 100190, China.}\\[0.4cm]
{\tt E-mail:
{$^\star$baixj@sogang.ac.kr,
$^\dagger$huyp@nuaa.edu.cn,
$^\ddagger$bhl@sogang.ac.kr,
$^\S$zhangyl@itp.ac.cn}}}
}





\begin{abstract}
The holographic charged fluid with anomalous current in Einstein-Maxwell gravity has been generalized from the infinite boundary to the finite cutoff surface by using the gravity/fluid correspondence. After perturbing the boosted Reissner-Nordstrom (RN)-AdS black brane solution of the Einstein-Maxwell gravity with the Chern-Simons term, we obtain the first order perturbative gravitational and Maxwell solutions, and calculate the stress tensor and charged current of the dual fluid at finite cutoff surfaces which contains undetermined parameters after demanding regularity condition at the future horizon. We adopt the Dirichlet boundary condition and impose the Landau frame to fix these parameters, finally obtain the dependence of transport coefficients in the dual stress tensor and charged current on the arbitrary radical cutoff $r_c$. We find that the dual fluid is not conformal, but it has vanishing bulk viscosity, and the shear viscosity to entropy density ratio is universally $1/4\pi$. Other transport coefficients of the dual current turns out to be cutoff-dependent. In particular, the chiral vortical conductivity expressed in terms of thermodynamic quantities takes the same form as that of the dual fluid at the asymptotic AdS boundary, and the chiral magnetic conductivity receives a cutoff-dependent correction which vanishes at the infinite boundary.

PACS number: 11.25.Tq,~04.65.+e,~04.70.Dy

\end{abstract}

\maketitle

\vspace*{1.cm}

\newpage

\section{Introduction}

The AdS/CFT correspondence~\cite{Maldacena:1997re,Gubser:1998bc,Witten:1998qj,Aharony:1999ti} provides a remarkable connection between a gravitational theory and a quantum field theory. According to the correspondence, the gravitational theory in an asymptotically AdS spacetime can be formulated in terms of a quantum field theory on its boundary. In particular, the dynamics of a classical gravitational theory in the bulk is mapped to a strongly coupled quantum field theory on the boundary. Therefore, AdS/CFT provides a useful tool and some insights to investigate the strongly coupled field theory from the dual classical gravitational theory~\cite{Herzog:2009xv,CasalderreySolana:2011us}.

The gravity/fluid correspondence can be considered as the long wave-length limit of the AdS/CFT correspondence, since the hydrodynamics is an effective description of an interacting quantum field theory in the long wave-length limit, i.e. when the length scales under consideration are much larger than the correlation length of the quantum field theory. The big advantage of the gravity/fluid correspondence in~\cite{Bhattacharyya:2008jc}  is that it provides a systematic way to map the boundary fluid to the bulk gravity, since it can construct the stress-energy tensor of the fluid order by order in a derivative expansion from the bulk gravity solution, and the shear viscosity $\eta$ and entropy density $s$ can be calculated from the first order stress-energy tensor \cite{Hur:2008tq, Erdmenger:2008rm, Banerjee:2008th, Son:2009tf, Tan:2009yg, Torabian:2009qk, Hu:2010sn}.
Besides the stress-energy tensor, this correspondence can also construct the dual conserved charged current if the Maxwell field is introduced in the bulk gravity, The thermal conductivity and electrical conductivity of the boundary fluid can be extracted from the conserved charged  current \cite{Hur:2008tq, Erdmenger:2008rm, Banerjee:2008th}.
Furthermore, it has been shown that {the chiral magnetic effect (CME)} and  the chiral vortical effect (CVE) can be brought into the hydrodynamics after adding the Chern-Simons term of Maxwell field in the action~\cite{Son:2009tf, Tan:2009yg, Hu:2011ze, Kalaydzhyan:2011vx, Amado:2011zx}. The effect of Chern-Simons term was first discussed in three dimensions where the Maxwell theory becomes massive due to the introduction of the Chern-Simons term~\cite{Deser:1982vy}. In addition, the Chern-Simons term can also affect the phase transition of Holographic Superconductors in four dimensions~\cite{Tallarita:2010vu} and stability of the Reissner-Nordstrom (RN) black holes in AdS space in five dimension \cite{Nakamura:2009tf}.

{
Usually, the dual field theory in AdS/CFT correspondence resides in the boundary with infinite radial coordinate. According to the renormalization group (RG) viewpoint, the radial direction of the bulk spacetime corresponds to energy scale of the dual field theory~\cite{Balasubramanian:1998de,de Boer:1999xf,Susskind:1998dq}. The energy scale on the infinite boundary is the UV fixed point value, hence it can not be reached by experiments.} {Recently, many works have been devoted to discuss the dual physics at the finite cutoff surface, and the RG flow was investigated in several approaches, such as 
holographic Wilsonian renormalization group~\cite{Heemskerk:2010hk,Faulkner:2010jy} and sliding membrane~\cite{Iqbal:2008by,Bredberg:2010ky}, and it has been found that those apparently different approaches are actually equivalent \cite{Sin:2011yh}. The physics at finite cutoff surface $r=r_c$ which implies finite energy scale becomes important, and the dependence of transport coefficient of dual fluid on the cutoff surface $r_c$ is usually interpreted as the corresponding renormalization group (RG) flow.}
{
On the other hand, it was shown in ~\cite{Bredberg:2010ky,Bredberg:2011jq,Compere:2011dx} that a given solution of the incompressible Navier-Stokes(NS) equations could be mapped to a unique solution of the vacuum Einstein equations. The non-relativistic fluid could live on the cutoff surface $r=r_c$ in Rindler space-time implies a deep relationship between the fluid and gravity. Using the Dirichlet boundary condition at the boundary and regularity at the future horizon, this relation is generalized to the non-relativistic fluid dual to asymptotically AdS gravity at a finite cutoff surface~\cite{Cai:2011xv,Niu:2011gu}. The relativistic fluid can also be investigated on the cutoff surface~\cite{Compere:2012mt,Kuperstein:2011fn}, and interestingly, the holographic cutoff fluid in the asymptotically AdS gravity can be re-expressed as an emergent hypersurface fluid which is non-conformal but has the same value of the shear viscosity as the infinity boundary fluid. Imposing Dirichlet-type boundary conditions on the hypersurface amounts to a constructive definition of the hypersurface theory,  whose underlying physics is yet to be understood.
The RG flow of the fluid takes the form of field redefinitions of the boundary hydrodynamic variables~\cite{Kuperstein:2011fn}, which is different from the usual Holographic RG approach that fixing observables at the infinity boundary and writing down a bulk-to-cutoff map that preserves those observables.}


In this paper, 
by following the spirit of the gravity/fluid correspondence
 \cite{Bredberg:2010ky,Cai:2011xv,Kuperstein:2011fn}.
We investigate the holographic charged fluid at the finite cutoff surface in asymptotically AdS gravity, which contains the Chern-Simons term of the Maxwell field in the bulk. Like the infinite boundary case~\cite{Hur:2008tq},
after perturbating the boosted RN-AdS black brane solution of the Einstein-Maxwell gravity, we obtain the first order perturbative gravitational and Maxwell solutions, and calculate the stress-energy tensor and charged current of the dual fluid to first order on the finite cutoff surface. We find that they contain undetermined parameters related to the boundary conditions and gauge conditions, and we explicitly express the dependence of the dual stress tensor and charged current on these parameters. A little difference from the case with infinite boundary, here we adopt the Dirichlet boundary condition and Landau frame to fix these parameters.

The dual fluid on the hypersurface is found to be non-conformal but has vanishing bulk viscosity, and the sheer viscosity takes the same value as boundary fluid. The familiar thermodynamic relation still holds for arbitrary cutoffs. These results are consistent with previous results of \cite{Kuperstein:2011fn}. We also obtain the chiral vortical conductivity and chiral magnetic conductivity for the dual fluid on the hypersurface. The former is the same as that found in \cite{Son:2009tf}, but the later receives a cutoff dependent correction
\begin{eqnarray}\label{correction}
\delta \tsigma_B = -\frac{c~g Q}{r_c^2\sqrt{3f(r_c)}},
\end{eqnarray}
where $c$ is the anomaly coefficient, $f(r_c)=1-\frac{2M^4}{r_c^4}+\frac{Q^2}{r_c^6}$, $Q$ and $M$ are the charge and mass of RN-AdS black brane and related to temperature and chemical potential of the dual fluid, see \cite{Son:2009tf} for explicit relations.

{In order to understand this correction, let's first recall that in the quantum field theory with anomalies \cite{Bardeen:1984pm}, the consistent current is defined to be the variation of the generating functional with respect to the gauge field. It is gauge covariant up to an additive gauge non-covariant term known as the Bardeen-Zumino (BZ) polynomial. One may subtract the BZ polynomial from the consistent current and obtain the gauge-covariant current.
The consistent current arises more naturally in the context of AdS/CFT correspondence, defined as
\begin{equation}\label{current}
\tJ^\mu = r_{c}^4\frac{1}{\sqrt{-\gamma}} \frac{\delta S_{cl}}{\delta \tA_\mu} = J^{\mu}+\delta \tJ^{\mu}~,
\end{equation}
where $J^{\mu}$ is the covariant current which produces the same chircal magnetic conductivity as that of ref. \cite{Son:2009tf} and $\delta\tJ^{\mu}$ is the Bardeen-Zumino polynomial 
\begin{equation}
P^{\mu}_{BZ}=\delta \tJ^{\mu}=\frac{c}{6}\epsilon^{\mu\nu\rho\lambda} \tA_{\nu} \tF_{\rho\lambda},
\end{equation}
which comes from the variation of the Chern-Simons term in the AdS/CFT context.}
For vanishing background field, this term doesn't contribute. However, the hypersurface theory naturally endows with a background field
\begin{equation}
\tA_{\mu}={ g\frac{\sqrt 3 Q}{r_c^2 \sqrt{f(r_c)}}u_{\mu}},
\end{equation}
which eventually gives rise to the correction in chiral magnetic conductivity. In particular, we should note that $\tA_{\mu}$ is not a constant field since $Q$ and $M$ are both local functions of cutoff surface coordinates $x^{\mu}$. A similar situation for the boundary fluid where the correction arises from a constant background field has been discussed in \cite{Amado:2011zx}.

The rest of the paper is organized as follows. In Sec.~II, we present a quick review of basics of hydrodynamics with anomalies. In Sec.~III we briefly review some properties of the RN black brane solution. In Sec.~IV, we construct the perturbative solution to the first order. In Sec.~V, we extract the dual stress-energy tensor and the conserved current from this first order perturbation solution. In Sec.~VI, we use the Dirichlet boundary condition and Landau frame to fix the undetermined parameters, and obtain the dependence of transport coefficients on the radial cutoff $r_c$. Sec.~VII is devoted to the conclusion and discussion.

\section{hydrodynamics with anomalies}
{ Any interacting quantum field theory admits a low energy effective description in terms of fluid dynamics. All the physical information of fluid is contained in the stress tensor $\tau^{\mu \nu}$ and charge currents $J^{\mu}_{I}$, where $I$ enumerates all the conserved charges in the system. They can be constructed order by order in the derivative expansion. In a (3+1)-dimensional hydrodynamics with triangle anomalies, in the presence of an external field, the energy-momentum tensor and charged current satisfy \cite{Son:2009tf}
\begin{eqnarray}
\partial_{\mu} \tau^{\mu\nu}=F^{\nu\mu} J_{\mu},~~~~
\partial_{\mu} J^{\mu}=c E^{\mu}B_\mu.
\end{eqnarray}
Note that the charge current is gauge covariant here. For a relativistic fluid, the heat transfer inevitably involves mass or momentum transfer, therefore the definition of fluid velocity field $u^{\mu}$ needs a careful consideration. It's conventional to introduce the Landau frame  $u_{\mu}\tau^{\mu\nu}=-\rho u^\nu$ and $u_{\mu}J^{\mu}=-n$ to fix this ambiguity. In this frame, the most general form of constitutive relations up to the first order gradient expansion are
\begin{eqnarray}
\tau^{\mu\nu}&=&\rho u^\mu u^\nu+p P^{\mu\nu}-2\eta\sigma^{\mu\nu}-\zeta\theta P^{\mu\nu},\label{Tmn} \\
J^\mu&=&n u^\mu+ \nu^{\mu},\\
\nu^{\mu}&=&-\kappa P^{\mu \nu }\partial _{\nu}(\frac{\mu }{T})+\sigma_{E}E^{\mu }+\sigma _{B}B^{\mu }+\xi \omega ^{\mu },\label{Jm}
\end{eqnarray}
where $P^{\mu\nu}=g^{\mu\nu}+u^{\mu}u^{\nu}$ and the first-order gradient expansion tensors are expressed as
\begin{eqnarray}
\sigma^{\mu\nu}& = &
\frac{1}{2}P^{\mu\alpha}P^{\nu\beta}\(\partial_{\alpha}u_{\beta}+\partial_{\beta}u_{\alpha}-\theta
P_{\alpha\beta}\),~
\theta=\partial_{\mu}u^{\mu},\\
E^{\mu}& = &F^{\mu\nu}u_{\nu},~~~B^\mu=\frac{1}{2}\epsilon^{\mu\nu\alpha\beta}u_{\nu}F_{\alpha\beta},
~~~\omega^{\mu}=\frac{1}{2}\epsilon^{\mu\nu\alpha\beta}u_{\nu}\partial_{\alpha}u_{\beta}.
\end{eqnarray}

In the case of conformal fluid, stress tensor has to be traceless, which requires the energy density $\rho = 3p$ where $p$ is pressure, and the bulk viscosity $\zeta$ has to vanish.

Another important concept is the entropy current which takes following form
\begin{equation}
s^{\mu} = s u^{\mu} -\dfrac{\mu}{T}\nu^{\mu}+ D \omega^{\mu} + D_B B^{\mu}
\end{equation}
in the hydrodynamics with anomalies. The physical requirement for entropy current is $\partial_\mu s^\mu\geq 0$ which guarantees local entropy production and is consistent with second thermodynamics law. This basically reflects the generic fact that the system starting an non-equilibrium state evolves towards an equilibrium state. With this physical requirement, imposing Landau frame and using the thermodynamic relation $\rho+p=Ts+n \mu$, the chiral magnetic conductivity $\sigma_B$ and chiral vortical conductivity $\xi$ can be fixed in term of anomaly coefficient $c$ and thermodynamic quantities
\begin{eqnarray}
\sigma_B =c\left(\mu-\frac{1}{2}\frac{n\mu^2}{\rho+p}\right),~~~
\xi = {c}\left(\mu^2-\frac{2}{3}\frac{n\mu^3}{\rho+p}\right).
\end{eqnarray}
One can also obtain the same results from certain generic properties of equilibrium partition function, see \cite{Banerjee:2012iz}.}

\section{Action and black brane solution}
The action of the 5-dimensional Einstein-Maxwell gravity with Chern-Simons term can be
\begin{eqnarray}
\label{action1} S & = & \frac{1}{16 \pi \GN}\int_\mathcal{M}~d^5x \sqrt{-g_{(5)}} \left(R-2 \Lambda \right)\nn\\
& &-\frac{1}{4g^2}\int_\mathcal{M}~d^5x \sqrt{-g_{(5)}}(F^2+\frac{4\kappa_{cs} }{3}\epsilon ^{LABCD}A_{L}F_{AB}F_{CD}),
\end{eqnarray}
and the equations of motion are
\begin{eqnarray}
\label{eqs1}
R_{AB } -\frac{1}{2}Rg_{AB}+\Lambda g_{AB}-\frac{1}{2g^2}\left(F_{A C}{F_{B }}^{C}-\frac{1}{4}g_{AB}F^2\right)&=&0~, \\
\nabla_{B} {F^{B}}_{A}-\kappa_{cs} \epsilon_{ABCDE}F^{BC}F^{DE} &=&0,\nonumber~~
\end{eqnarray}
where $\Lambda = -6/\ell^2$, we set $16 \pi \GN =1$ and $\ell=1$ for later convenience.
In our paper, the black brane solution we are interested in is the $5$-dimensional charged RN-AdS black brane solution
\cite{Cai:1998vy,Cvetic:2001bk,Anninos:2008sj}
\begin{eqnarray}
ds^2=-r^2f(r) dt^2+\frac{dr^2}{r^2f(r)}+r^2(dx^2 +dy^2 +dz^2),\label{Solution}
\end{eqnarray}%
where
\begin{eqnarray}
\label{f-BH}
f(r) &=& 1-\frac{2M}{r^{4}}+\frac{Q^2}{r^6},~~F = -g\frac{2\sqrt 3 Q}{r^3}dt \wedge dr. ~~
\end{eqnarray}%
Note that, the above RN-AdS black brane solution (\ref{Solution}) is still the solution of  (\ref{eqs1}), although the Chern-Simons term affects the Maxwell equation. From~(\ref{Solution}), we can easily find that the outer horizon of the black brane is located at $r=r_{+}$, where $r_{+}$ is the largest root of $f(r)=0$, and its Hawking temperature and entropy density are
\begin{eqnarray}
T_{+}&=&\frac{(r^2f(r))'}{4 \pi}|_{r=r_{+}}=\frac{1}{2 \pi r_{+}^3}(4M-\frac{3Q^2}{r_{+}^2}),\label{Temperature}\\
s&=&\frac{r_{+}^3}{4\GN }.\label{entr}
\end{eqnarray}
In addition, this black brane solution (\ref{Solution}) rewritten in the Eddington-Finkelstin coordinate system is
\begin{eqnarray}\label{Solution1}
ds^2 &=& - r^2 f(r)dv^2 + 2 dv dr + r^2(dx^2 +dy^2 +dz^2),\\
F &=& -g\frac{2\sqrt 3 Q}{r^3}dv \wedge dr. \notag~~
\end{eqnarray}
where $v=t+r_*$ and $r_*$ is the tortoise coordinate related to radial coordinate $r$ by $dr_*=dr/(r^2f)$.
The coordinate singularity is avoided in this coordinate system.

Note that, since we will consider the holographic charged fluid at some cutoff hypersurface $r=r_c$ (here $r_c$ is a constant), thus we can first make a coordinate transformation $v\rightarrow v/\sqrt{f(r_c)}$ on the above solution (\ref{Solution1}) before the boost transformation. The motivation of this coordinate transformation is to make the induced metric on the cutoff surface explicitly conformal to flat metric $ds^2=r_{c}^2 (-dv^2+dx^2+dy^2 +dz^2)$. It should be emphasized that the Hawking temperature in this new coordinates system is $T=T_{+}/\sqrt{f(r_c)}$, and the RN-AdS black brane solution in the new coordinates system is
\begin{eqnarray}\label{Solution2}
ds^2 &=& - \frac{r^2 f(r)}{f(r_c)}dv^2 + \frac{2}{\sqrt{f(r_c)}} dv dr + r^2(dx^2 +dy^2 +dz^2),\\
F &=& -g\frac{2\sqrt 3 Q}{r^3 \sqrt{f(r_c)}}dv \wedge dr. \notag~~
\end{eqnarray}
We boost the RN-AdS black brane solution to acquire nonzero velocities
\begin{eqnarray}   \label{rnboost}
ds^2 &=& - \frac{r^2 f(r)}{f(r_c)}( u_\mu dx^\mu )^2 - \frac{2}{\sqrt{f(r_c)}} u_\mu dx^\mu dr + r^2 P_{\mu \nu} dx^\mu dx^\nu, \\
A&=&\frac{\sqrt 3 g Q}{r^2 \sqrt{f(r_c)}}u_{\mu}dx^{\mu}, \label{Externalfield}~~
\end{eqnarray}
with
\begin{equation}
u^v = \frac{1}{ \sqrt{1 - \beta_i^2} },~~u^i = \frac{\beta_i}{ \sqrt{1 - \beta_i^2} },~~P_{\mu \nu}= \eta_{\mu\nu} + u_\mu u_\nu, \label{velocity}
\end{equation}
where velocities $\beta^i $, $M$ and $Q$ are constants, $x^\mu=(v,x_{i})$ are the cutoff surface coordinates, $P_{\mu\nu}$ is the projector onto spatial directions, and the indices in the boundary are raised and lowered with the Minkowski metric
$\eta_{\mu\nu}=\text{diag}\{-,+,+,+\}$.
Note that (\ref{Externalfield}) naturally gives rise to a background field for the dual fluid on cutoff surfaces. Unlike the boundary fluid case, an additional non-dynamical external field is not needed to obtain the electric conductivity and chiral magnetic conductivity.

\section{The first order perturbative solution}
In this section, we will consider the region between the outer horizon and cutoff surface
$r_{+}\leq r \leq r_c$,
and then perturb the boosted black brane~(\ref{rnboost}) to make its velocity and temperature nonuniform. Therefore, we can extract the transport coefficients
of the dual fluid from this perturbative solution via the Gravity/Fluid correspondence. According to the Gravity/Fluid correspondence, the perturbation is proceeded in the following.
Firstly, we promote the above constant parameters $\beta^i $, $M$, $Q$ to be slowly-varying functions of the
cutoff boundary coordinates $x^\mu=(v,x_{i})$. Therefore, the metric~(\ref{rnboost}) and Maxwell field (\ref{Externalfield}) will no longer be the solutions of the equations of motion~(\ref{eqs1}), and we need to add extra gravitational and Maxwell fields to make the equations of motion satisfied again. Before discussing these extra fields in detail, we define the following useful tensors
\begin{eqnarray}
&&W_{IJ} = R_{IJ} + 4g_{IJ}+\frac{1}{2g^2}\left(F_{IK}{F^{K}}_J +\frac{1}{6}g_{IJ}F^2\right),\label{Tensors1}\\
&&W_A = \nabla_{B} {F^{B}}_{A}-\kappa_{cs} \epsilon_{ABCDE}F^{BC}F^{DE}~~,\label{Tensors2}
\end{eqnarray}
where we use the convention $\epsilon_{vxyzr}=+\sqrt{-g}$. Obviously, when we take the parameters $\beta^i $, $M$ and $Q$ as functions of $x^\mu$ in~(\ref{rnboost}),
$W_{IJ}$ and $W_{A}$ will be nonzero and can be proportional to the derivatives of
the parameters. Usually, these nonzero terms $-W_{IJ}$ and $-W_{A}$ are considered as the source terms $S_{IJ}$
and $S_{A}$. Therefore, the extra gravitational and Maxwell fields are added into (\ref{rnboost}) and (\ref{Externalfield}) such that they can deduce some correction terms to cancel the source terms and make the equations of motion still satisfied. In this work, we only consider the first order perturbative case. For this first order case, the parameters are expanded around $x^\mu=0$ to the first order
\begin{eqnarray}
\beta_i&=&\beta^i(0)+\partial_{\mu} \beta_{i}|_{x^\mu=0} x^{\mu},~~M=M(0)+\partial_{\mu}
 M|_{x^\mu=0} x^{\mu},~~~Q=Q(0)+\partial_{\mu} Q|_{x^\mu=0} x^{\mu}. \label{Expand}
\end{eqnarray}
And we can assume $\beta^i(0)=0$ because it is always possible to choose coordinates to set $u^{\mu}=(1,0,0,0)$ at any given point $x^{\mu}$~\cite{Bhattacharyya:2008jc}.
After inserting the metric (\ref{rnboost}), (\ref{Externalfield}) and (\ref{Expand}) into $W_{IJ }$ and $W_{A}$, the nonzero $-W_{IJ }$, $-W_{A}$ can be considered as the first order source terms $S^{(1)}_{IJ }$ and $S^{(1)}_{A}$. Therefore, after fixing some gauge (the `background field' gauge in \cite{Bhattacharyya:2008jc}, $G$ represents the perturbed metric with corrections)
\begin{equation}
G_{rr}=0,~~G_{r\mu}\propto u_{\mu},~~Tr((G^{(0)})^{-1}G^{(1)})=0,\label{gauge}
\end{equation}
and considering the spatial $SO(3)$ symmetry preserved in the background metric~(\ref{Solution1}), the choice for the first order extra gravitational and Maxwell fields around $x^\mu=0$ can be
\begin{eqnarray}\label{correction}
&&{ds^2_{(1)}} = \frac{ k(r)}{r^2}dv^2 + 2\frac{h(r)}{\sqrt{f(r_c)}}dv dr + 2 \frac{j_i(r)}{r^2}dv dx^i
 +r^2 \left(\alpha_{ij} -\frac{2}{3} h(r)\delta_{ij}\right)dx^i dx^j, \\
&&A^{(1)} = a_v (r) dv + a_i (r)dx^i~~.\label{correction1}
\end{eqnarray}
Note that the gauge $a_r(r)=0$ has been chosen.
Therefore, the first order perturbation solution can be obtained
 from the vanishing $W_{IJ} = (\text{effect from correction})
  - S^{(1)}_{IJ }$ and $W_{A} = (\text{effect from correction}) - S^{(1)}_{A}$.
  Here, the ``effect from correction'' means the correction to $W_{IJ}$ and $W_{A}$ from (\ref{correction}) and (\ref{correction1}).

For the first order gravitational equations, they are complicated which have been listed in the appendix~\ref{A}. For the first order Maxwell equations, they are
\begin{eqnarray}\label{Maxwell}
&&W_v = \frac{f(r)}{r}\left\{ r^3 {a_v}' (r) -  \frac{4\sqrt 3 g Q}{\sqrt{f(r_c)}}h(r) \right\}' - S^{(1)}_{v}(r)=0~,\\\nn
&& W_r = - \frac{1}{r^3}\left\{  r^3\sqrt{f(r_c)} {a_v}'(r) - 4\sqrt 3 g Q h(r)  \right\}' - S^{(1)}_r (r)=0~,\\\nn
&&W_i = \frac{1}{r} \left\{ r^3 f(r) {a_i}'(r) + \frac{2 \sqrt 3 g Q
\sqrt{f(r_c)} }{r^4} j_i (r) \right\}' -  S^{(1)}_i (r)=0~,\nn
\end{eqnarray}
where
\begin{eqnarray}
&&S_v^{(1)}(r)=-g\frac{2 \sqrt{3}}{r^3} \left(\partial _vQ+Q \partial _i\beta _i\right),\\\nn
&&S_r^{(1)}(r)=0,\\\nn
&&S_i^{(1)}(r)=g \left(\frac{\sqrt{3} Q \partial _v\beta _i}{r^3}
+\frac{\sqrt{3} Q \partial _iM}{r^3 r_c^4 f\left(r_c\right)}
+\frac{\sqrt{3} \partial _xQ \left(-2 M+r_c^4\right)}{r^3 r_c^4 f\left(r_c\right)}\right)-\kappa _{\text{cs}} \frac{48 g^2 Q^2}{r^6 \sqrt{f\left(r_c\right)}}\epsilon^{i j k}\partial _j\beta _k~,
\end{eqnarray}
and prime means derivative with respect to $r$ coordinate. Note that the cutoff effect has been incorporated in these equations through their dependence on $r_c$. In addition, from the above first order gravitational and Maxwell equations, there are several interesting relations between these equations
\begin{eqnarray}
 &&W_v + \dfrac{r^2 f(r)}{\sqrt{f(r_c)}}W_r =0 ~:~ S_v^{(1)} + \dfrac{r^2 f(r)}{\sqrt{f(r_c)}}S_r^{(1)} =0, \notag \\
 &&W_{vv} + \dfrac{r^2 f(r)}{\sqrt{f(r_c)}}W_{vr} =0 ~:~ S_{vv}^{(1)} + \dfrac{r^2 f(r)}{\sqrt{f(r_c)}}S_{vr}^{(1)} = 0, \notag \\
 &&W_{vi} + \dfrac{r^2 f(r)}{\sqrt{f(r_c)}} W_{ri} =0 ~:~ S_{vi}^{(1)} +\dfrac{r^2 f(r)}{\sqrt{f(r_c)}}S_{ri}^{(1)} = 0, \label{constraint}
 \end{eqnarray}
which can be considered as the constraint equations. In our paper, after using the first order source terms, we can further rewrite these constrain equations~(\ref{constraint}) as
\begin{eqnarray}
&&\partial _vQ+Q \partial _i\beta _i=0,  \\\nn
&&3 \partial _vM+4 M \partial _i\beta _i=0,\\\nn
&& \partial _iM+4 M \partial _v\beta _i=\frac{4 M \left(Q \partial _iQ-r_c^2\partial _iM \right)}{f\left(r_c\right) r_c^6}~~.
\end{eqnarray}
Later, we can see that there are some underlying physical interpretations for these equations, i.e., these equations are related to the conservation equations of the zeroth order stress-energy tensor and conserved current.

Therefore, after adding the correction terms, the first-order metric expanded in boundary derivatives around $x^{\mu}=0$ can be explicitly given as
\begin{eqnarray}
ds^{2}&=&\frac{2}{\sqrt{f(r_c)}}dvdr-\frac{r^{2}}{f(r_c)}f(M_{0},Q_{0},r)dv^{2}+r^{2}dx_{i}^{2}-\frac{r^{2}}{f(r_c)}x^{\mu }C_{1}(r)\partial _{\mu }Mdv^{2}-2x^{\mu }\partial _{\mu }\beta_{i}dx^{i}dr \notag\\
& &-\frac{2}{\sqrt{f(r_c)}}x^{\mu}r^{2}(1-f(M_{0},Q_{0},r))\partial _{\mu }\beta _{i}dx^{i}dv-\frac{r^{2}}{f(r_c)}x^{\mu }C_{2}(r)\partial _{\mu }Qdv^{2}+\frac{ k(r)}{r^2}dv^2 \notag\\
&&+ 2\frac{h(r)}{\sqrt{f(r_c)}}dv dr + 2 \frac{j_i(r)}{r^2}dv dx^i
 +2r^2 \left(\sigma_{ij} -\frac{1}{3} h(r)\delta_{ij}\right)dx^i dx^j,\label{GloballySolution1}
\end{eqnarray}
where
\begin{eqnarray}
f(M_{0},Q_{0},r)&=&f(M(x^{\mu}),Q(x^{\mu}),r)|_{x^{\mu}=0},~~C_{1}(r)=\frac{\partial f(M(x^{\mu}),Q(x^{\mu}),r)}{\partial M}|_{x^{\mu}=0},\notag\\
C_{2}(r)&=&\frac{\partial f(M(x^{\mu}),Q(x^{\mu}),r)}{\partial Q}|_{x^{\mu}=0},~~\sigma_{ij}=\partial_{(i} \beta_{j)}-\frac{1}{3} \delta_{ij}\partial_k \beta^k.
\end{eqnarray}
And the global first-order metric defined on the whole cutoff surface can be constructed by replacing (\ref{GloballySolution1}) in a covariant form, i.e., replacing $\partial_i\beta^i$ as $\partial _{\lambda }u^{\lambda }$ in (\ref{GloballySolution1})~\cite{Bhattacharyya:2008jc}.

\section{The Stress tensor and charged current}
The information of the dual fluid like its stress tensor and charged current can be directly extracted from the above global first-order perturbative solution. In addition, we can also first extract the dual stress tensor and charged current from (\ref{GloballySolution1}), and then rewrite the dual stress tensor and charged current in a covariant form. Here we use the later calculation.

According to the Gravity/Fluid correspondence, the dual stress tensor $\tau _{\mu\nu}$ can be obtained through the following relation ~\cite{Myers:1999psa}
\begin{eqnarray}\label{relation}
\label{Tik-CFT} \sqrt{-h}h^{\mu\nu}\langle\tau _{\nu\sigma}\rangle=\lim_{r\rightarrow
r_c }\sqrt{-\gamma }\gamma ^{\mu\nu}T_{\nu\sigma},
\end{eqnarray}
where $h_{\mu\nu}$ is the background metric upon which the dual field theory resides, $\gamma ^{\mu\nu}$ is the boundary metric on the cutoff surface obtained from the well-known ADM decomposition
\begin{eqnarray}
ds^2 = \gamma_{\mu\nu}(dx^\mu + V^\mu dr)(dx^\nu + V^\nu dr) + N^2 dr^2~.
\end{eqnarray}
We impose the Dirichlet-type boundary condition on the cutoff surface to fix the induced metric $\gamma_{\mu\nu}$. In particular, all the unknown functions in the first order correction metric ${ds^2_{(1)}}$ and gauge field $A^{(1)}$ vanish at $r_c$. It's easy to obtain the induced metric $\gamma_{\mu\nu} = r_c^2\eta_{\mu\nu}$, where $\eta_{\mu\nu}$ is flat spacetime metric.  Using (\ref{relation}) the expectation value of the
first order stress tensor of the dual fluid $\tau _{\mu \nu}$ is
\begin{equation}
\tau_{\mu \nu}=r_c^2T_{\mu \nu}. \label{StressTensor}
\end{equation}
$T_{\mu\nu}$ is the boundary stress tensor which is defined through~\cite{Balasubramanian:1999re,Emparan:1999pm,Mann:1999pc}
{
\begin{equation}
 T_{\mu\nu}\equiv \dfrac{-2}{\sqrt{-\gamma}}\dfrac{\delta S_{cl}} {\delta\gamma^{\mu\nu}} = 2\left(K_{\mu\nu}-K\gamma _{\mu\nu}
-C\gamma _{\mu\nu}\right)\,. \label{TabCFT}
\end{equation}}
The extrinsic curvature is $K^{\mu\nu}=-\frac{1}{2}(\nabla^{\mu}n^{\nu}+\nabla^{\nu}n^{\mu})$ and $n^{\mu}$ is the normal vector of the constant hypersurface $r=r_c$ pointing toward the $r$ increasing direction. {
Note that the third term in (\ref{TabCFT}), $C\gamma_{\mu\nu}$, is needed
to cancel the divergence of the boundary stress tensor when the cutoff $r_c$ is taken to infinity. Although the constant $C$ can be fixed by certain regularity conditions in the asymptotic boundary case, for instance,
$C=3$ in asymptotic $AdS_5$ case, it remains to be a free parameter on the cutoff surface.
We shall keep its presence throughout the paper. Actually, it only changes the fluid pressure and energy density by a constant and doesn't affect any transport coefficients. In the ref.\cite{Kuperstein:2011fn}, it's taken to be the same value as the asymptotic case.}

The stress tensor of dual fluid on cutoff surface can be explicitly expressed from (\ref{GloballySolution1}) by using (\ref{TabCFT})
\beqs
    \tau^{(0)}_{v v} &=& 2\left(C-3 \sqrt{f\left(r_c\right)}\right) r_c^4,\nn\\
    \tau^{(0)}_{i i} &=&\frac{-4 M+2\left(3-C \sqrt{f\left(r_c\right)}\right) r_c^4}{\sqrt{f\left(r_c\right)}},\nn\\
    \tau^{(1)}_{v v} &=&-2\partial _i\beta _i r_c^3+2\sqrt{f\left(r_c\right)} r_c^5 h'\left(r_c\right),
    \\
    \tau^{(1)}_{v i} &=& -\frac{Q \partial _iQ}{ f\left(r_c\right) r_c^3}+\frac{\partial _iM}{f\left(r_c\right) r_c}-\partial _v\beta _i r_c^3-\sqrt{f\left(r_c\right)} j_i'\left(r_c\right) r_c,\nn\\
    \tau^{(1)}_{i j} &=&2\left(2 \delta _{i j}\partial _k\beta _k-\partial _{(i}\beta _{j)} \right)r_c^3+2\delta _{i j}\frac{\partial _k\beta _k \left(2 M-3 r_c^4\right)}{3 f\left(r_c\right) r_c}
    -\sqrt{f\left(r_c\right)} r_c^5 \alpha_{i j}'\left(r_c\right), \nn\\
    & & -2\delta _{i j}\left(
    \frac{2}{3} \sqrt{f\left(r_c\right)} r_c^5 h'\left(r_c\right)
    +\frac{1}{2}\sqrt{f\left(r_c\right)}r_c k'\left(r_c\right)\right).\nn
\eeqs

The consistent current of the dual fluid is defined as
\begin{equation}\label{current}
\tJ^\mu = r_{c}^4\frac{1}{\sqrt{-\gamma}} \frac{\delta S_{cl}}{\delta \tA_\mu} = -r_{c}^4 \frac{N}{g^2} (\tF^{r \mu}+\frac{4\kappa_{cs} }{3}\epsilon ^{r\mu \rho \sigma \tau }\tA_{\rho}\tF_{\sigma \tau })~,
\end{equation}
where $\tA_\mu$ is the gauge field projected to boundary. Note the second term comes from the variation of Chern-Simons term in the action and appears to be gauge non-covariant, but from the dual fluid point of view, it's allowed due to the presence of anomaly. It leads to a cutoff dependent correction in the chiral magnetic coefficient. Explicitly, the components of dual current are given by
{
\beqs
&&\tJ^{\nu}_{(1)}=
\frac{\sqrt{f\left(r_c\right)} r_c^3 a_v'\left(r_c\right)}{g^2},\\
&&\tJ_{(1)}^{i}=-\frac{2 \sqrt{3} Q j_i\left(r_+\right)}{g r_+^4}-\frac{\sqrt{3} Q \partial _v\beta _i}{g r_+ \sqrt{f\left(r_c\right)}}-\frac{\sqrt{3} Q \partial _iM}{g r_+ r_c^4 f\left(r_c\right){}^{3/2}}\nn\\
&&~~~~~~~~~-\frac{\sqrt{3} \partial _iQ \left(-Q^2+r_c^6 f\left(r_c\right)\right)}{g r_+ r_c^6 f\left(r_c\right){}^{3/2}}-\kappa _{\text{cs}} \frac{4 Q^2  \left(r_+^4-3 r_c^4\right)}{r_+^4 r_c^4 f\left(r_c\right)}\epsilon^{i j k}\partial _j\beta _k,
\eeqs
}
where $j_x(r_+)/r_+^4$ is a little complicated, and it has been listed in the (\ref{ji(r+)}) in appendix~\ref{C}.

\section{Thermodynamics and transport coefficients}

{Now let's solve the bulk equations $W_{I}=0$ and $W_{IJ}=0$ by imposing some suitable conditions and study the thermodynamics and transport properties of the dual fluid on cutoff surface. We can start by integrating out equation $W_{rr}=0$ to obtain $h(r)$ and solve $a_v(r)$ from $W_{r}=0$ and $k(r)$ from $W_{vv}=0$,
\begin{eqnarray}
h(r)&=&{\Chtwo}+\frac{{\Chone}}{r^4}\nn\\
a_v(r)&=&{\Cavtwo}+\frac{\Cavone}{r^2}-\frac{2 {\Chone} g Q}{\sqrt{3} r^6 \sqrt{f\left(r_c\right)}}\\
k(r)&=&{\Cktwo}+{\Ckone} r^2-\frac{2 {\Chtwo} r^4}{f\left(r_c\right)}+\frac{4 {\Chone} \left(-Q^2+M r^2\right)}{3 r^6 f\left(r_c\right)}+\frac{2 \Cavone Q}{\sqrt{3} g r^2 \sqrt{f\left(r_c\right)}}+\frac{2 r^3 \partial _i\beta _i}{3 \sqrt{f\left(r_c\right)}}\nn.
\end{eqnarray}
Then we impose the regularity condition at the future horizon and $\alpha_{ij}(r_c)=0$ to obtain $\alpha_{ij}(r)$ from (\ref{A8})
\begin{eqnarray}\label{alpha_ij}
\alpha_{ij}(r)&=&\alpha(r)\left\{(\partial_i \beta_j + \partial_j \beta_i )-\frac{2}{3} \delta_{ij}\partial_k \beta^k \right\}.
\end{eqnarray}
where $\alpha(r)$ is
\begin{eqnarray}
\alpha(r)&&= \int_{r_c }^{r}\frac{s^{3}-r_{+}^3}{-s^{5}f(s)}ds.
\end{eqnarray}
Note that $\Ckone$ should vanish for the solution of (\ref{A6}) being consistent with (\ref{alpha_ij}), which can be checked by inserting $h(r)$, $a_v(r)$ and $ \alpha_{ii}(r)$ into (\ref{A6}). Next, we can fix $\Chone,$ $\Chtwo,$ $\Cktwo$, $ \Cavone$ and $ \Cavtwo$ by requiring $h(r_c)=0$, $a_v(r_c)=0$, $k(r_c)=0$ and another two conditions from imposing Landau frame $J_{(1)}^{v}=0$ and $ \tau^{(1)}_{v v}=0$, see Appendix \ref{B}. The equations involving $j_i(r)$ and $a_i(r)$ are coupled to each other. The detailed solving procedure is given in the Appendix \ref{C}. The integration constant for $a_i(r)$ is fixed by imposing $a_i(r_c)$=0. The solution of $j_i(r)$ contains two constants $\CTHREE$ and $\CFOUR$, which can be fixed by $\tau^{(1)}_{v x}=0$ (Landau frame condition) and $j_i(r_c) = 0$.}
The results are summarized as follows
\beqs
    & &{\Chone}=-\frac{\partial _i\beta _i r_c^3}{4 \sqrt{f\left(r_c\right)}},\quad{\Chtwo}=\frac{\partial _i\beta _i}{4 \sqrt{f\left(r_c\right)} r_c},\quad{\Ckone}=0,\nn\\
    & &\Cavone=-\frac{\sqrt{3} g Q \partial _i\beta _i}{2 f\left(r_c\right) r_c},\quad{\Cavtwo}=\frac{g Q \partial _i\beta _i}{\sqrt{3} f\left(r_c\right) r_c^3},\quad{\Cktwo}=-\frac{\partial _i\beta _i \left(-10 M+r_c^4\right)}{6 f\left(r_c\right){}^{3/2} r_c},\nn\\
    & &{\Cthree}=0,\quad{\Cfour}=\partial _v\beta _x r_c^3 \left(-1+\sqrt{f\left(r_c\right)}\right)+\frac{Q \partial _xQ-\partial _xM r_c^2}{r_c^3 \sqrt{f\left(r_c\right)}}.
\eeqs

Therefore, the non-zero components of boundary fluid energy stress tensor are
\beqs
    & &\tau^{(0)}_{v v} = 2\left(C- 3\sqrt{f\left(r_c\right)} \right) r_c^4, \quad \tau^{(0)}_{i i} =\frac{-4 M+2\left(3- C\sqrt{f\left(r_c\right)}\right) r_c^4}{\sqrt{f\left(r_c\right)}},\nn\\
   & &\tau^{(1)}_{i j} = - 2r_+^3\sigma _{ij},
\eeqs
which can be further rewritten in a covariant form
\begin{eqnarray}
\tau_{\mu \nu}=\rho\,u^\mu u^\nu+p P^{\mu\nu}-2\eta\sigma^{\mu\nu}-\zeta\theta P^{\mu\nu}, \label{StressTensor1}
\end{eqnarray}
where the energy density $\rho$, pressure $p$, shear viscosity $\eta$ and bulk viscosity $\zeta$ are
\begin{eqnarray}
\rho=2\left(C-3 \sqrt{f\left(r_c\right)} \right) r_c^4,~~~p=\frac{-4 M+2\left(3- C\sqrt{f\left(r_c\right)}\right) r_c^4}{\sqrt{f\left(r_c\right)}},
~~~\eta=r_+^3,~~~\zeta=0. \label{ets}
\end{eqnarray}
Note that although the bulk viscosity is vanishing in this case, the dual fluid on the cutoff surface is still not conformal, which is consistent with the result in \cite{Kuperstein:2011fn}.

The chemical potential is defined as
\begin{eqnarray}
\mu =  A_v (r_c)-A_v (r_+)  ~.\label{chemical potential}
\end{eqnarray}
Following the discussion in \cite{Hur:2008tq},  we can find that its first order expression is
\begin{eqnarray}
\mu = \frac{\sqrt{3}g Q}{\sqrt{f\left(r_c\right)}}\left( \frac{1}{r_+^2}-\frac{1}{r_c{}^2}\right)~,
\end{eqnarray}
which keeps the same expression but here $M$, $Q$ and $r_+$ are not constants.
The entropy density $s$ of dual fluid can be computed through
\begin{equation}
s=\left(\frac{\partial p}{\partial T}\right)_{\mu}=4\pi r_{+}^3. \label{entr1}
\end{equation}
which is consistent with the entropy density of the black brane solution in (\ref{entr}). It is convenient to check this equation if we express $p$ in the functions of $r_+$ and $ Q$. The shear viscosity to entropy ratio is universally $1/4\pi$.
Furthermore, the familiar thermodynamic relation still holds on arbitrary cutoff surface
\begin{equation}
\rho+p-sT=n\mu\,, \label{Thermorelation}
\end{equation}
where $T$ is the temperature of the dual fluid related to the Hawking temperature of the black brane solution by $T=\frac{T_+}{\sqrt{f(r_c)}}$  and $n$ is particle number density defined in (\ref{zeroth order current}).
Thus we can conclude that the thermodynamical properties of the dual charged fluid
 are universal on the cutoff surface.

The zeroth and first order charged current of the dual fluid are
\beqs
\tJ_{(0)}^{\mu} &=& \frac{2\sqrt 3 Q}{g} u^\mu =: n u^\mu\,,\label{zeroth order current}\\
\tJ_{(1)}^{\mu } &=&-\kappa P^{\mu \nu }\partial _{\nu}(\frac{\mu }{T}%
)+\sigma _{E}E^{\mu }+\tsigma _{B}B^{\mu }+\xi \omega ^{\mu }, \label{first order current}
\eeqs
where $n$ is particle number density and
\beqs
E^{\mu}&=& \tF^{\mu\nu }u_{\nu},
~B^{\mu }=\frac{1}{2}\epsilon ^{\mu \nu \rho \sigma}u_{\nu}\tildeF_{\rho \sigma},
~\omega ^{\mu }=\frac{1}{2}\epsilon ^{\mu \nu \rho \sigma }u_{\nu}\partial _{\rho}u_{\sigma}\,.
\eeqs
Here $\tF_{\mu \nu}$ is defined at the cutoff surface $r=r_c$ through $\tA_{\mu}= g\frac{\sqrt 3  Q}{r_c^2 \sqrt{f(r_c)}}u_{\mu}$, electric and magnetic fields are given by
\beqs
E^i  =  \frac{\sqrt{3} g Q \partial _iM}{4 M r_c^2 \sqrt{f\left(r_c\right)}}-\frac{\sqrt{3} g \partial _iQ}{r_c^2 \sqrt{f\left(r_c\right)}}~,~~
B^i  =  -\frac{1}{2} \epsilon^{ijk}\left( \frac{2\sqrt{3} g Q \partial _j\beta _k}{r_c^2 \sqrt{f\left(r_c\right)}}\right),~~\omega^i = -\frac{1}{2}\epsilon^{ijk}\partial_j\beta_k~.
\eeqs
The transport coefficients are found to be
\beqs
\kappa &=&\frac{16 \pi ^2 r_+^7 T_+^3}{g^2 r_c^{10} \sqrt{f\left(r_c\right)} f'(r_c)^2}\quad,~
 \sigma_E = \frac{16 \pi ^2 r_+^7 T_+^2}{g^2 r_c^{10} f'\left(r_c\right)^2},\nn\\
\tsigma_B &=& -\frac{8 Q \left(3 r_c^2-2 r_+^2\right) \kappa _{\text{cs}}}{\sqrt{3} g r_+^2 r_c^2 \sqrt{f\left(r_c\right)}}+\frac{24 \sqrt{3} Q^3 \left(r_c^2-r_+^2\right){}^2 \kappa _{\text{cs}}}{g r_+^4 r_c^9 \sqrt{f\left(r_c\right)} f'\left(r_c\right)},\nn\\
\xi &=& -\frac{24 Q^2 \left(r_c^2-r_+^2\right){}^2 \kappa _{\text{cs}}}{r_+^4 r_c^4 f\left(r_c\right)}+\frac{96 Q^4 \left(r_c^2-r_+^2\right){}^3 \kappa _{\text{cs}}}{r_+^6 r_c^{11} f\left(r_c\right) f'\left(r_c\right)}\,.
\eeqs
In the large $r_c$ limit, they can be expanded in the power series of $r_c$
\beqs
    \kappa & = &\frac{\pi ^2 r_+^7 T_+^3}{4 g^2 M^2}+\frac{3 \pi ^2 Q^2 r_+^7 T_+^3}{8 g^2 M^3 r_c^2} +\mathcal{O}(r_c^{-4})\,,\nn\\
    \sigma_E & = &\frac{\pi ^2 r_+^7 T_+^2}{4 g^2 M^2}+\frac{3 \pi ^2 Q^2 r_+^7 T_+^2}{8 g^2 M^3 r_c^2}+\mathcal{O}(r_c^{-4})\,,\nn\\
    \tsigma_B & =&-\frac{\sqrt{3} Q \left(2 M+3 r_+^4\right) \kappa _{\text{cs}}}{g M r_+^2}+\frac{Q \left(28 M^2-36 M r_+^4+27 r_+^8\right) \kappa _{\text{cs}}}{4 \sqrt{3} g M^2 r_c^2}+\mathcal{O}(r_c^{-4})\,,\nn\\
    \xi & = & -\frac{12 Q^2 \kappa _{\text{cs}}}{M}+\frac{6 Q^2 \left(4 M^2+3 r_+^8\right) \kappa _{\text{cs}}}{2 M^2 r_+^2 r_c^2}+\mathcal{O}(r_c^{-4}).~\label{TransportCoeff}
\eeqs
It's easy to see that these transport coefficients recover the previous results of boundary fluid \cite{Hur:2008tq,Son:2009tf}, if the cutoff is taken to be infinity.

In addition, the thermal conductivity and electrical conductivity happen to satisfy the Wiedermann-Franz law at arbitrary cutoff surface~\cite{Hur:2008tq,Hu:2010sn}
\begin{equation}
\kappa=\sigma_E T.
\end{equation}

It is illuminating to rewrite the chiral magnetic conductivity $\sigma_B $ and the chiral vortical conductivity $\xi$ in terms of thermodynamic quantities
\beqs
\tsigma_B & = &c\left(\mu-\frac{1}{2}\frac{n\mu^2}{\rho+p}\right)-\frac{c~g Q}{r_c^2\sqrt{3f(r_c)}},\label{sigmaB}\\
\xi &=& c\left(\mu^2-\frac{2}{3}\frac{n\mu^3}{\rho+p}\right), \label{sigmaxi}
\eeqs
where the anomaly coefficient $c=-{8 \kappa _{\text{cs}}}/{g^2}$.
In the asymptotic boundary, 
we can recover the result in \cite{Son:2009tf,Amado:2011zx}.
Now, let's analyze the origin of the second term in (\ref{sigmaB}). Note that, the dual charged current in (\ref{current}) actually can be decomposed into two parts, the gauge covariant current $J^\mu$ and the correction gauge dependent current $\delta \tJ^\mu$ coming from the variation of Chern-Simons term
\begin{eqnarray}\label{current1}
J^\mu&\equiv&\tJ^\mu-\delta \tJ^\mu=-r_{c}^4 \frac{N}{g^2}\tF^{r \mu},\label{Jmem}\\
\delta \tJ^\mu&\equiv&- r_{c}^4 \frac{N}{g^2} \frac{4\kappa_{cs} }{3}\epsilon ^{r\mu \rho \sigma \tau }\tA_{\rho}\tF_{\sigma \tau }= \frac{c}{6}\epsilon ^{\mu \rho \sigma \tau }\tA_{\rho}\tF_{\sigma \tau }.\label{Jkappa}
\end{eqnarray}
And at the finite cutoff surface,
the general conservation equations could be obtained from the Gauss-Codazzi relations of Einstein equations (\ref{Tensors1}) and constraint relations of the Maxwell equations in (\ref{Tensors2}). 
In our case, the constraint equations at the cutoff surface turns out to be
\begin{eqnarray}\label{conservation}
\partial _{\mu }\tau^{\mu  \nu }&=&\tF^{ \nu \mu }J_{\mu }=\tF^{ \nu \mu }\(\tJ_{\mu }-\frac{c}{6}\epsilon ^{\mu \rho \sigma \tau }\tA_{\rho}\tF_{\sigma \tau }\),~~~\\
\partial _{\mu }J^{\mu } &=& c\, {E}^{\mu}{B}_{\mu},~~~~~~
\partial _{\mu }\tJ^{\mu } = \frac{c}{3}\, {E}^{\mu}{B}_{\mu},\label{Anomalous}
\end{eqnarray}
which gives the anomalous Ward identities for the stress tensor and current~\cite{Banerjee:2012iz}.
At the zeroth order, it could be check that they reduce to
\beqs
& &\partial _{\mu }\tau_{(0)}^{\mu  \nu }=\tildeF^{ \nu \mu }J^{(0)}_{\mu },
~~~~\partial _{\mu }J_{(0)}^{\mu } = 0,
\eeqs
which are just the constraint equations in~(\ref{constraint}).

As $\delta \tJ^\mu$ in (\ref{Jkappa}) only contributes to the chiral magnetic term in the current,
and after some calculation, we find that the chiral magnetic conductivity $\sigma_B$ can be divided into two parts,
\begin{eqnarray}
\sigma_B\equiv c\left(\mu-\frac{1}{2}\frac{n\mu^2}{\rho+p}\right),\qq
\delta \tsigma_B\equiv-\frac{c~g Q}{r_c^2\sqrt{3f(r_c)}},
\end{eqnarray}
where $\sigma_B$  is contribution from the gauge covariant current $J^{\mu }$,
and $\delta \tsigma_B$ is contribute from the current $\delta \tJ^{\mu }$.
The coefficient $\sigma_B$  is consistent with the relation in \cite{Son:2009tf}
because there the current is gauge covariant.
 $\delta \tsigma_B$ come from the gauge dependent corrections,
a similar situation for the boundary fluid where the correction arises from a
constant background field has been discussed in~\cite{Amado:2011zx}.
We can also conclude that our result from holographic calculation is consistent with the result
from the the equilibrium partition functions in \cite{Banerjee:2012iz}.



\section{Conclusion and discussion}

In this paper, we generalize the dual charged fluid on the infinite boundary case to the finite cutoff surface case via the gravity/fluid correspondence. Like the same procedure as the infinite boundary case, we first lift the parameters of the boosted RN black brane in the Einstein-Maxwell gravity with Chern-Simons term to functions of finite cutoff boundary coordinates, then solve for the corresponding correction terms and obtain the first order perturbative gravitational and Maxwell solutions, based on which we calculate the stress tensor and current for the dual fluid by using the Fluid/Gravity correspondence. The stress tensor and charged current on the finite cutoff surface depend on undetermined parameters which are related to the boundary and gauge conditions. We explicitly express the dependence of the dual stress tensor and charged current on these parameters. Different choices of these parameters could correspond different dual physics on the finite cutoff surface. In particular, we use the Dirichlet boundary condition, regularity at the future horizon and choose Landau frame to fix these parameters. Eventually, we work out the explicit dependence of transport coefficients in the dual stress tensor and charged current on the radial cutoff $r_c$.

We find that the hypersurface fluid is non-conformal but has vanishing bulk viscosity. The parity-preserving sector, i.e.~shear viscosity and thermodynamic relation, remains the same as boundary fluid. Other transport coefficients of the dual current are found to be cutoff dependent.
It is interesting to pursue whether the dependence of transport coefficients of dual fluid on the cutoff surface $r_c$ obtained here could be identified with the RG flow of these coefficients.

Another interesting observation is that the chiral vortical conductivity remain the same relation with the thermodynamic quantities as the boundary fluid, but
there appears a discrepancy in the chiral magnetic conductivity. This discrepancy can be traced back to the appearance of a gauge non-covariant term $\delta \tJ$ in the definition of the consistent current.  $\delta \tJ$ corresponds to the Bardeen-Zumino polynomial in the quantum field theory with anomalies.
In the presence of a background gauge field $\tA$, it gives rise to a correction in the chiral magnetic conductivity.
At the asymptotic AdS boundary, the background gauge field goes to zero, so does the correction term. If one adopts the covariant current, this discrepancy would disappear entirely. This distinction should be recognized as a general feature of the quantum field theory with anomalies and does not have anything to do with the AdS/CFT duality , Holography RG flow, cutoff surface etc.


\section{Acknowledgements}

We thank Prof. Rong-Gen Cai, Chan-Yong Park, Dr. Zhang-Yu Nie, Peng Sun and Jian-Hui Zhang for useful discussions. Our special thanks go to Dr.~Jyotirmoy Bhattacharya, Yang Zhou and  the referee for their valuable comments regarding the revision of this work. Dr.~Ya-Peng Hu thanks Center for Quantum Spacetime (CQUeST) in Sogang University for the kind hospitality during his visit. Y.L Zhang is grateful for Prof.~Rong-Gen Cai's encouragement and support. He would like to also thank the ``2012 Asia Pacific School/Workshop on Cosmology and Gravitation" held at Yukawa Institute for Theoretical Physics (YITP-W-11-26), for their hospitality and financial support.
This work is supported by China Postdoctoral Science Foundation under Grant No.20110490227 and National Natural Science Foundation of China (NSFC) under grant No.11105004, and by the National Research Foundation of Korea(NRF) grant funded the Korea government(MEST) through the Center for Quantum Spacetime(CQUeST) of Sogang University with grant number 2005-0049409, and partially by grants from NSFC (No. 10821504, No. 10975168 and No. 11035008) and the Ministry of Science and Technology of China under Grant No. 2010CB833004.

\appendix

\section{The tensor components of $W_{\mu \nu }$ and $S_{\mu \nu }$}
\label{A} The tensor components of $W_{\mu \nu } = (\text{effect
from correction}) - S_{\mu \nu }$ are
\begin{eqnarray}
& &W_{vv} =-\frac{8 r^2 f(r) h(r)}{f\left(r_c\right)}+\frac{2 \left(2 Q^2-2 M r^2-r^6\right) f(r) h'(r)}{r^3 f\left(r_c\right)}-\frac{4 Q f(r) a_v'(r)}{\sqrt{3} g r \sqrt{f\left(r_c\right)}}\\
& &\qquad+\frac{f(r) k'(r)}{2 r}-\frac{1}{2} f(r) k''(r)-S_{vv}^{(1)}~,\nn\\
& &W_{vi}=-\frac{\sqrt{3} Q f(r) a_x'(r)}{g r \sqrt{f\left(r_c\right)}}+\frac{3 f(r) j_x'(r)}{2 r}-\frac{1}{2} f(r) j_x''(r)-S_{vi}^{(1)}(r)~,\\
& &W_{vr}=\frac{4 Q a_v'(r)}{\sqrt{3} g r^3}+\frac{8 h(r)}{\sqrt{f\left(r_c\right)}}+\frac{2 \left(-2 Q^2+2 M r^2+r^6\right) h'(r)}{r^5 \sqrt{f\left(r_c\right)}}\nn\\
& &\qquad-\frac{\sqrt{f\left(r_c\right)} k'(r)}{2 r^3}+\frac{\sqrt{f\left(r_c\right)} k''(r)}{2 r^2}-S_{vr}^{(1)}\\
& &W_{ri} =\frac{\sqrt{3} Q a_i'(r)}{g r^3}-\frac{3 \sqrt{f\left(r_c\right)} j_i'(r)}{2 r^3}+\frac{\sqrt{f\left(r_c\right)} j_i''(r)}{2 r^2}-S_{ri}^{(1)}\\
& &W_{rr} = \frac{5 h'(r)}{r}+h''(r)-S_{rr}^{(1)}\\
& &W_{ii}=8 r^2 h(r)+\frac{\left(5 Q^2-14 M r^2+11 r^6\right) h'(r)}{3 r^3}+\frac{1}{3} r^4 f(r) h''(r)+\frac{f\left(r_c\right) k'(r)}{r}\nn\\
& &\qquad-\frac{2 Q \sqrt{f\left(r_c\right)} a_v'(r)}{\sqrt{3} g r}+\frac{\left(Q^2+2 M r^2-5 r^6\right) \alpha _{i i}'(r)}{2 r^3}-\frac{1}{2} r^4 f(r) \alpha _{i i}''(r)-S_{ii}^{(1)} \label{A6}\\
& &W_{ij}=\frac{\left(Q^2+2 M r^2-5 r^6\right) \alpha _{i j}'(r)}{2 r^3}-\frac{1}{2} r^4 f(r) \alpha _{i j}''(r)-S_{ij}^{(1)},~(i\neq j)\\
& &W_{ij}-\dfrac{1}{3}\delta_{ij}\left(\sum_k W_{kk}\right)=\frac{\left(Q^2+2 M r^2-5 r^6\right) \alpha _{i j}'(r)}{2 r^3}-\frac{1}{2} r^4 f(r) \alpha _{i j}''(r)-S_{ij}^{(1)} \label{A8}
\end{eqnarray}
where the first order source terms are
\begin{eqnarray}
S_{vv}^{(1)}(r)&=&-\frac{3\partial _vM}{r^3 \sqrt{f\left(r_c\right)}}+\frac{3 Q \partial _vQ}{r^5 \sqrt{f\left(r_c\right)}}-\frac{\left(-2 Q^2+2 M r^2+r^6\right) \partial _i\beta _i}{r^5 \sqrt{f\left(r_c\right)}}\\
S_{vi}^{(1)}(r)&=&
-\frac{Q \left(3 Q^2+2 M r^2+3 r^6\right) \partial _iQ}{2 r^5 r_c^6 f\left(r_c\right){}^{3/2}}+\frac{\left(3 Q^2+2 M r^2+3 r^6\right) \partial _iM}{2 r^5 r_c^4 f\left(r_c\right){}^{3/2}}\nn\\
& &+\frac{\partial _iM}{r^3 \sqrt{f\left(r_c\right)}}+\frac{\left(3 Q^2+2 M r^2+3 r^6\right) \partial _v\beta _i}{2 r^5 \sqrt{f\left(r_c\right)}}\\
S_{vr}^{(1)}(r)&=&\frac{\partial _i\beta _i}{r}\\
S_{ri}^{(1)}(r)&=&-\frac{3 \partial _v\beta _i}{2 r}+\frac{3 Q \partial _iQ}{2 r r_c^6 f\left(r_c\right)}-\frac{3 \partial _iM}{2 r r_c^4 f\left(r_c\right)}\\
S_{rr}^{(1)}(r) &=&0, \\
S_{ij}^{(1)}(r) &=&\left(\delta _{i j}\partial _k\beta _k+3\partial _{(i}\beta _{j)}\right)r\sqrt{f(r_c)}.
\end{eqnarray}

\section{The explicit expression of $\tau^{(1)}_{v v}$, $\tau^{(1)}_{v x}$ and  $J_{(1)}^{v}$} \label{B}
\beqs
    \tau^{(1)}_{v v} &=&-\frac{4 \partial _i\beta _i r_c \left(C-3 \sqrt{f\left(r_c\right)}\right)}{3 \sqrt{f\left(r_c\right)}}+{\Ckone} \left(-2C+9 \sqrt{f\left(r_c\right)}\right)+\frac{{\Cktwo} \left(-2 C+9 \sqrt{f\left(r_c\right)}\right)}{ r_c^2}\nn\\
    & &+\frac{2{\Chone} \left(2 C \left(Q^2+r_c^6 \left(-1+f\left(r_c\right)\right)\right)-3 \sqrt{f\left(r_c\right)} \left(3 Q^2+r_c^6 \left(-3+4 f\left(r_c\right)\right)\right)\right)}{3 r_c^8 f\left(r_c\right)}\\
    & &+\frac{2{\Chtwo} r_c^2 \left(2 C+3 \left(-3+f\left(r_c\right)\right) \sqrt{f\left(r_c\right)}\right)}{f\left(r_c\right)}+\frac{2\Cavone Q \left(-2 C+9 \sqrt{f\left(r_c\right)}\right)}{\sqrt{3} g r_c^4 \sqrt{f\left(r_c\right)}}\nn\\
\tau^{(1)}_{v x}&=&\frac{{\Cfour}}{ \sqrt{f\left(r_c\right)}}+{2\Cthree}~ r_c^4 \left(-C+3 \sqrt{f\left(r_c\right)}\right) f\left(r_c\right)\nn\\
    & &+\frac{-Q \partial _xQ+r_c^2 \left(\partial _xM-\partial _v\beta _x r_c^4 \left(-\sqrt{f\left(r_c\right)}+f\left(r_c\right)\right)\right)}{ r_c^3 f\left(r_c\right)}\\
J_{(1)}^{v} &=& \frac{2 Q \partial _i\beta _i}{\sqrt{3} g r_c \sqrt{f\left(r_c\right)}}+\frac{\sqrt{3} {\Ckone} Q}{g r_c^2}+\frac{\sqrt{3} {\Cktwo} Q}{g r_c^4}+\frac{2 {\Chone} Q \left(-2 Q^2+2 M r_c^2+3 r_c^6 f\left(r_c\right)\right)}{\sqrt{3} g r_c^{10} f\left(r_c\right)}\nn\\
& &-\frac{2 \sqrt{3} {\Chtwo} Q \left(1+f\left(r_c\right)\right)}{g f\left(r_c\right)}-\frac{2 \Cavone \left(-Q^2+r_c^6 f\left(r_c\right)\right)}{g^2 r_c^6 \sqrt{f\left(r_c\right)}}~~.
\eeqs

\section{The exact form of $j_i(r)$ and $a_i(r)$} \label{C}
For the $a_{i}(r)$ and $j_{i}(r)$, we can solve them from equations $W_i =0$ and $W_{ri} =0$. However, these equations are more difficult to solve since $a_{i}(r)$ and $j_{i}(r)$ are coupled to each other. These equations $W_i =0$ and $W_{ri} =0$ explicitly are
\begin{eqnarray}
&& \left(r^3 f(r) a_i'(r)+\frac{2 \sqrt{3} g Q}{r^4} \sqrt{f(r_c)} j_i(r)\right)'=r S_i^{(1)}(r)~, \label{B1}\\
&&\frac{\sqrt{3} Q a_i'(r)}{g r^3}-\frac{3 \sqrt{f\left(r_c\right)} j_i'(r)}{2 r^3}+\frac{\sqrt{f\left(r_c\right)} j_i''(r)}{2 r^2}=S_{r i}^{(1)}(r)~. \label{B2}
\end{eqnarray}
Since we just consider the first order case, we ignore the up indexes $(1)$ in $S_{r
i}^{(1)}(r)$ and $S_i^{(1)}(r)$ for the convenience in the following.

Integrating eq. (\ref{B1}) from the horizon $r_+$ to $r$, we get
\begin{eqnarray}
r^3 f(r) a_i'(r)+2 \sqrt{3} g Q \left(\frac{j_i(r)}{r^4}-\frac{j_i\left(r_+\right)}{r_+^4}\right)=\int_{r_+}^r ds\, s~S_i(s),
\end{eqnarray}
and imposing the boundary condition that $a_i(r)$ vanish at cutoff surface
\begin{eqnarray}
a_i(r) &=& \int_{r_c}^r dw
\frac{1}{w^3f(w)}\left(\int_{r_+}^w ds\,s\,S_{i}(s)-2\sqrt{3}g Q(\frac{j_{i}(w)}{w^4}-\frac{j_i(r_+)}{r_{+}^4})\right).
\end{eqnarray}

After some algebra, eq. (\ref{B2}) is reduced to
\begin{eqnarray}
j_i{}^{\prime\prime }(r)-\frac{3}{r} j_i'(r)-\frac{12 Q^2} {r^8 f(r)}j_i(r)  = \zeta_i(r),
\label{appeqj}
\end{eqnarray}
where
\begin{equation}
\zeta_i(r)=-\frac{12 Q^2}{r^4 f(r)} \frac{ j_i\left(r_+\right)}{r_+^4}+\dfrac{2 r^2}{\sqrt{f(r_c)}} S_{r i}(r)-\frac{2 \sqrt{3} Q}{g r^4 f(r)\sqrt{f(r_c)}} \int _{r_+}^rds\, s\,S_i(s).
\end{equation}

And then we can write out a particular solution to (\ref{appeqj})
\begin{eqnarray}
j_P(r)=b_1(r) j_{H_1}(r)+b_2(r) j_{H_2}(r),
\end{eqnarray}
where
\begin{eqnarray}
b_1(r)&=&-\int _r^{r_c }dx \, \frac{j_{H_2}(s) \zeta _i(s)}{s^3}, \\
b_2(r)&=&r^3 \partial _v\beta _i+\int
_r^{r_c }ds\, \left(\frac{j_{H_1}(s) \zeta _i(s)}{s^3}+3 s^2 \partial _v\beta_i\right).
\end{eqnarray}
and
\begin{eqnarray}
j_{H_1}(r)&=&r^4 f(r), \\
j_{H_2}(r)&=&r^4 f(r) \int _r^{r_c}\frac{ds}{s^5 f(s)^2
}.
\end{eqnarray}
are two linearly independent homogeneous solutions of (\ref{appeqj}). Here, the $3 s^2 \partial _v\beta _i$ term is added to cancel the divergence of the integral.
With the above formulas, the general solution to (\ref{appeqj}) can be represent as
\begin{eqnarray}
j_i(r)=j_P(r)+\CTHREE~j_{H_1}(r)+\CFOUR~j_{H_2}(r).
\end{eqnarray}
In summary,
\begin{eqnarray}
j_i(r)&=&-r^4 f(r)\int _r^{r_c }ds\, s f(s) \zeta _i(s) \int _s^{r_c }\frac{dw}{w^5 f(w)^2} +r^4 f(r) \left(\int _r^{r_c }\frac{ds}{s^5 f(s)^2}\right)\nonumber\\
&& \Big(r^3 \partial _v\beta _i +\int _r^{r_c}ds \left(s f(s) \zeta _i(s)+3 s^2 \partial _v\beta _i\right)\Big)+\CTHREE j_{H_1}(r)+\CFOUR j_{H_2}(r). \label{appjsol}
\end{eqnarray}

To obtain $j_i(r_+)$, we take $r\to r_+$ limit to (\ref{appjsol})
and get
\begin{eqnarray}\label{j at rp}
\frac{j_i\left(r_+\right)}{r_+^4}&=&\frac{r_c^3\partial _v \beta _i+\int _{r_{+}}^{r_c}ds\,s f(s) \zeta _i(s)+\CFOUR}{r_{+}^5 f'(r_+)}
\end{eqnarray}
Noth that, $\zeta_{i}(r_{+})$ also contains $j_i(r_+)$, thus (\ref{j at rp}) is in fact the equation related to $j_i(r_+)$. After solving the above equation, we can obtain $j_i(r_+)$
\begin{eqnarray}
& &\frac{j_x\left(r_+\right)}{r_+^4}=\frac{{\Cfour}}{r_c^5 f'\left(r_c\right)}
+\frac{\partial _xM \left(-Q^2 \left(r_+^3-3 r_+^2 r_c+2 r_c^3\right)+r_+^2 r_c^2 \left(r_+^4 r_c-r_+ r_c^4+6 M \left(-r_++r_c\right)\right)\right)}{r_+^3 r_c^{12} f\left(r_c\right){}^{3/2} f'\left(r_c\right)}\nn\\
& &\quad+\frac{\partial _v\beta _x \left(-Q^2 \left(r_+^3-3 r_+^2 r_c+2 r_c^3\right)+r_+^2 r_c^2 \left(6 M \left(-r_++r_c\right)+r_+ r_c \left(r_+^3+r_c^3 \left(-1+\sqrt{f\left(r_c\right)}\right)\right)\right)\right)}{r_+^3 r_c^8 \sqrt{f\left(r_c\right)} f'\left(r_c\right)}\nn\\
& &\quad\frac{Q \partial _xQ \left(Q^2 \left(-r_+^3+r_c^3\right)-r_c^2 \left(-2 M \left(5 r_+^3-6 r_+^2 r_c+r_c^3\right)+r_c \left(r_+^6+r_+^3 r_c^3-3 r_+^2 r_c^4+r_c^6\right)\right)\right)}{r_+^3 r_c^{14} f\left(r_c\right){}^{3/2} f'\left(r_c\right)}\nn\\
& &\quad-\kappa _{\text{cs}} \left(\frac{4 \sqrt{3} g Q^3 \left(\partial _z\beta _y-\partial _y\beta _z\right) \left(r_+^6-3 r_+^2 r_c^4+2 r_c^6\right)}{r_+^6 r_c^{11} f\left(r_c\right) f'\left(r_c\right)}
\right)\label{ji(r+)}
\end{eqnarray}

In addition, there is a useful equation
\begin{eqnarray}
j'_i(r_c)=-\frac{r_c^2 \partial_v \beta _i}{f(r_c)}+\CTHREE(r^4f(r))'|_{r=r_c}-\frac{\CFOUR}{r_c f(r_c)}\,.
\end{eqnarray}


\end{document}